# A NOVEL OVERLAY IDS FOR WIRELESS SENSOR NETWORKS

Sumanta Saha, Md. Safiqul Islam, Md. Sakhawat Hossen
*School of Information and Communication Technology*
*The Royal Institute of Technology (KTH)*
*Stockholm, Sweden*

Mohammad Saiful Islam Mamun
*School of Computer and Systems Sciences*
*The Royal Institute of Technology (KTH)*
*Stockholm, Sweden*

## ABSTRACT

Wireless Sensor Network (WSN) consists of low cost sensor nodes which cannot afford to implement sophisticated security system in it. That is why intrusion detection architecture for WSN is considerably different and difficult to implement. Most of the current implementations are based on exchanging anomaly signals among the leaf level sensors resulting in too much power consumption. We propose a novel architecture for Intrusion Detection System (IDS) in WSN based on Hierarchical Overlay Design (HOD) that will distribute the overall responsibility of intrusion detection into entities and thus conserve memory and power of the nodes. The architecture uses layered design with GSM cell like structure based on special monitor nodes. The HOD structure enables the sensors to communicate using far less messages and thus conserve precious power and also saves memory by not implementing IDS module on each sensor. The proposal also uses rippling of alarm through layers and thus ensures proper delivery to the uppermost layer with redundancy.

## 1. INTRODUCTION

WSN, a subset of Wireless Ad hoc Network, is one of today's topics on demand. WSN consists of tiny sensors consisting of only the most required components for sensing the environment. The modules in the sensors are mainly battery powered and so are susceptible to power failure in case of over use. Due to this limitation, the sensors almost always tend to be limited in memory and processing power. On the other hand, most of the sensor systems are deployed in mission sensitive areas like war field, environmentally inaccessible terrains or disaster sites where the system needs to be protected against any illegal and malicious intrusion. But the inaccessibility of the area as well as the limitation of memory, power and processing makes the protection of WSN a challenging task.

Although there has been much advancement in the field of WSN architecture, little has been done on securing the system against intrusion. Existing proposals of IDS for wired systems cannot be implemented in WSN due to its low processing power and memory. Consequently, specialized IDS is necessary for WSN. In this paper, we have proposed a new architecture for IDS in WSN focusing to detect intrusion without losing much of valuable memory and power of the nodes. This architecture arranges the system of detection in hierarchical layers to reduce the overhead on the leaf level sensors and the overlay design ensures redundancy and quick response time in case of any intrusion.

## 2. CHALLENGES AND SECURITY THREATS IN WSN

All the security issues that are considered for wireless ad-hoc network should also be considered for wireless sensor network. However security mechanisms devised for wireless ad-hoc network (WAHN) cannot be applied to WSN because of the architectural inequality of these two networks. The common security vulnerabilities that cause attacks in WSN can be recognized:

- Lack of infrastructure
- Shared wireless medium
- Cooperative nature between the nodes
- Easy physical accessibility by the intruders
- Dynamic network topology
- Operational constraints

Various types of threats and attacks are now a day's mature topic in the area of WSN and they are well enumerated by the past researches in this particular field. A literature search on the topic of attacks on WSN, points to several studies carried out along possible remedies (Pathan, A.S.K. 2006), (Perrig, A. 2004).

## 3. INTRUSION DETECTION SYSTEM (IDS)

An intrusion is defined as any set of actions that attempt to compromise the integrity, confidentiality or availability of a resource in a host or network. According to the Network Security Bible – *"Intrusion detection and response is the task of monitoring systems for evidence of intrusions or inappropriate usage and responding to this evidence."* (Zhang, Y. and Lee, W. 2000).The basic idea of Intrusion Detection is observing user as well as program activities inside the system via auditing mechanism. Detection can be based on signature, e.g. searching network traffic for a series of bytes or packet sequences known to be malicious, or based on anomaly e.g. technique centered on the concept of a baseline for network behavior.

## 4. EXISTING IDS MODELS FOR WSN

Due to the key architectural differences between wired and wireless networks, their IDSs cannot be used interchangeably. There are different techniques for IDS in Wireless Sensor Network (WSN). Here we represent some existing IDS models for WSN.

- Intrusion detection by watching the communications of neighbourhood sensors (Roman, R. et al 2005)
- Decentralizing IDS to ensure redundancy (Da Silva, A. et al 2005)
- Applying non-cooperative Game Approach for IDS (Krishnan M. 2006)
- Steps to build an IDS using centralized routing method (Strikos. A. 2007)

## 5. PROPOSED MODEL

We propose a new model for intrusion detection which uses a hierarchical overlay design (HOD) based architecture. Our model concentrates on power saving of sensor nodes by distributing the responsibility of intrusion detection among different layers. The design is summarized below.

### 5.1 Intrusion Detection Entities

The total area of sensor network is divided into hexagonal regions (like GSM cells). Sensor nodes in each of the hexagonal area are monitored by a cluster node. Moreover, each cluster node is being monitored by a regional node. Fig 1 shows the general architecture of our model. In turn, Regional nodes will be controlled and monitored by the Base station. In this paper, we propose four layer architecture. Fig 2 shows our proposed four layer system architecture and relation among them. Following sections describe each of the layers in detail.

### 5.1.1 Sensor Nodes

The sensor nodes have two types of functionality. The first one is sensing and the second one is routing. Each of the sensor nodes will sense the environment and exchange data in between other sensor nodes and cluster node. Sensor nodes have limitation of memory, battery and computational power. To preserve these valuable resources, in our architecture, we do not have any intrusion detection modules installed in the leaf level sensor nodes.

### 5.1.2 Cluster Node

Cluster nodes are playing as a monitor node for the sensor nodes. One cluster node is assigned for each of the hexagonal area. It will receive the data from sensor nodes, analyze and aggregate the information and send it to regional node. It is more powerful than sensor nodes and has intrusion detection capability built into it. It works on different phases as shown in figure 3.

In data acquisition (Phase 1) it will receive the data from the sensor nodes and apply rules (Phase 2) to find misbehaving nodes and if any intrusion is detected (Phase 3) then it will generate an intrusion detection alert and send it to the regional node.

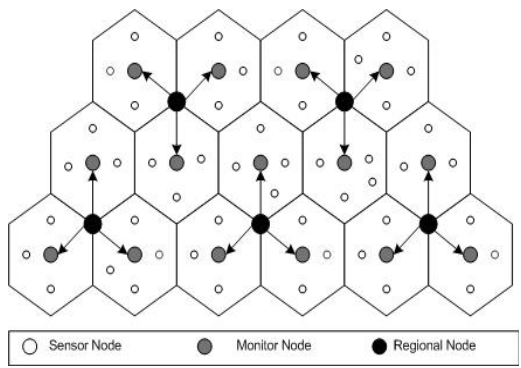

Figure 1. Hierarchical Overlay Design

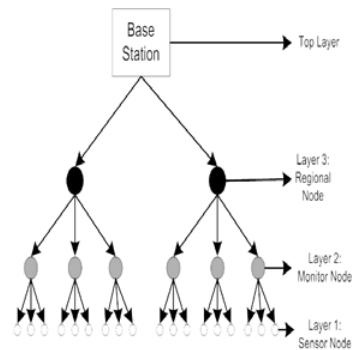

Figure 2. Four Layer Architecture

### 5.1.3 Regional Nodes

Regional nodes will monitor and receive the data from neighboring three cluster heads and send the aggregated alarm to the base station. It is also a monitor node like the cluster nodes with all the IDS functionalities.

The Inclusion of regional nodes in the architecture ensures a clear redundancy in the sense that if for some reason the cluster nodes are compromised regional nodes will be able to detect that and report that to the upper layer.

Another extra feature of the regional nodes is, as these nodes are communicating directly with the base station and the base station is obviously far out of range of the radio capacity of the regional nodes, they have to have a special communication module built inside for long range communication.

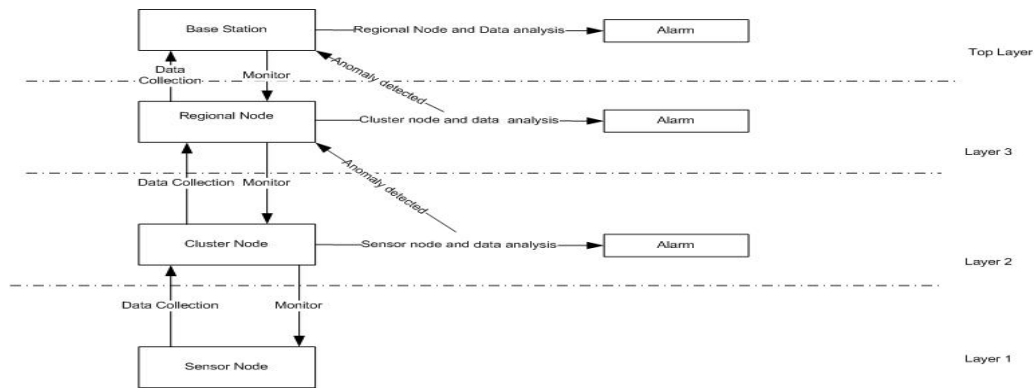

Figure 3. Intrusion Detection Architecture

### 5.1.4 Base Station

Base station is the topmost part of our architecture empowered with human support. It will receive the information from Regional nodes and distribute the information to the users based on their demand.

The base station is equipped with all the necessary instruments for reporting to the authority. It gathers all the information from the regional nodes and summarizes to generate a compact report of the intrusion in the network.

## 5.2 Intrusion Detection Mechanism

In our experiment we try to detect intrusion at multiple layers and we divide it into physical layer, MAC layer, network layer and application layer. If one layer fails to detect any anomaly then other layers will detect that. We use this multiple layers concept to make our IDS design robust.

### 5.2.1 Physical Layer:

Jamming is the primary physical layer attack. Identifying jamming attack can be done by the Received Signal Strength Indicator (RSSI) (Raymond, D. R. and Midkiff, S. F. 2008) (Bhuse, V. and Gupta, A. 2006), the average time required to sense an idle channel (carrier sense time), and the packet delivery ratio (PDR). In case of wireless medium, received signal strength has relation with the distance between nodes. Node tampering and destruction is another physical layer attack that can be prevented by placing nodes in secured area.

### 5.2.2 Mac Layer:

Link layer attacks are collision, denial of sleep and packet replay. We want to use S-MAC and Time Division Multiple Access (TDMA) to detect the anomaly at this layer.

a) TDMA (FZhang, Y. and Lee, 2000) is digital transmission process where each cluster node will assign different time slots for different sensor nodes in its region. During this slot every sensor node has access to the radio frequency channel without interference. If any attacker send packet using source address of any node, e.g. A, but that slot is not allocated to A then cluster node can easily detect that intrusion.

b)We can use S-Mac (*FZhang, Y. and Lee, 2000*) protocol to assign a wakeup and sleep time for the sensor nodes. As the sensor has limited power, we can use S-Mac for the energy conservation. If we receive any packet from source A while its sleeping period then the cluster node can easily detect the inconsistency.

### 5.2.3 Network Layer:

At this layer we can use route tracing to detect whether the packet really comes from the best route. If packet comes to the destination via different path rather than the desired path then the cluster node can detect possible intrusion.

### 5.3.4 Application Layer:

Our model suggests using three types watch dog. They are base station, regional node, cluster node. Sensor nodes will be monitored by upper layer watchdog cluster node and cluster nodes will be monitored by regional node watchdog and finally the top level watchdog base station will monitor the regional nodes. So, if any layer's node is compromised by the attacker then another layer's watch dog can easily detect the attack and generate alarm. Figure 3 illustrates a clear idea how intrusion detection system works.

## 6. CONCLUSION

Our design of IDS improves on other related designs in the way it distributes the total task of detecting intrusion using GSM cell-like architecture. It decouples the total work of intrusion detection into a four level hierarchy which results in a highly energy saving structure. Hierarchical watch dog concept acts as a basis of IDS where the top layer base station, cluster node and regional node are three hierarchical watchdogs. This paper proposes IDS in multiple layers to make the system architecture robust and redundant. Along with watchdog concept, the system also considers detecting intrusions in multiple OSI layers which makes the architecture even more attractive for a system like WSN.

## 7. FUTURE WORK

Although this work is purely on architecture of WSN and for that we did not concentrate on the IDS mechanisms to be used, but we have the plan to investigate the following areas in the future.

*a)* *Election procedure to select cluster and regional nodes:* Instead of choosing the cluster node and regional node manually, there will be an election process that will automatically detect the cluster node and regional node, thus making the system more rigid against natural calamity.

*b)* *Building a special simulator:* As all the previous researches were based on three layer architecture, we are planning to create our own simulator that will simulate the four layer OHD architecture of our system.